\documentclass[preprintnumbers]{revtex4}
\UseRawInputEncoding
\usepackage{amssymb}
\usepackage{amsmath}
\usepackage{graphicx}
\usepackage{dcolumn}
\usepackage{bm}
\usepackage{subfigure}
\usepackage{color}

\begin{document}

\title{Photon orbits and phase transition for Non-Linear charged Anti-de Sitter black holes}
\author{Yun-Zhi Du, Huai-Fan Li\footnote{the corresponding author}, Fang Liu, Li-Chun Zhang}
\affiliation{Department of Physics, Shan xi Da tong University, Da tong 037009, China\\
Institute of Theoretical Physics, Shan xi Da tong University, Da tong 037009, China}

\thanks{\emph{e-mail:duyzh13@lzu.edu.cn, huaifan999@163.com, sapphire513@126.com, zhlc2969@163.cn}}

\begin{abstract}
In this work, we investigate the relationship between the photon sphere radius and the first-order phase transition for the charged EPYM AdS black hole. Through the analysis, we find with a certain condition there exist the non-monotonic behaviors between the photon sphere radius, the impact parameter, the non-linear YM charge parameter, temperature, and pressure. And both the changes of photon sphere radius and impact parameter before and after phase transition can be regarded as the order parameter, their critical exponents near the critical point are equal to the same value $1/2$, just like the ordinary thermal systems. These indicate that there maybe exists a universal relationship of gravity nearby the critical point for a black hole thermodynamical system. Furthermore, the effect of impact parameter on the deflect angle is also investigated.
\end{abstract}

\maketitle

\section{Introduction}

Since the Hawking-Page phase transition \cite{Hawking1983} of an AdS spacetime was proposed by Hawking and Page in $1983$, which was explained to a confinement/deconfinement phase transition in gauge theory in Ref. \cite{Witten1998} and could also be understood as a solid/liquid phase transition \cite{Altamirano2013} by regarding the cosmological constant as pressure $P=-\frac{\Lambda}{8\pi}=\frac{(n-1)(n-2)}{16\pi l^2}$, black hole thermodynamics has attracted lots of attention. Subsequently the phase transition in the extended AdS/dS phase space has been widely considered \cite{Hendi2017a,Hennigar2017a,Frassin,Kubiznak2012,Cai2013,Ma2017,Mir2017b,Banerjee2017,
Simovic2019,Mbarek2019,Li2017a,Ma2018,Zhang2016,Dinsmore2020,
Cai1306,Caldarelli2000,Mann1207,Wei2015,Banerjee1109,Hendi1702,Bhattacharya2017,Zeng2017,Hendi1803,Zhang1502,
Cheng1603,Dolan2014,Altamirano2014,Du2021,Zhang2020}. Based on the similarity of thermodynamic behaviors for black holes and ordinary systems, the microstructure of black holes system was investigated by the Ruppeiner scalar curvature \cite{Wei2015} following the Ruppeiner geometry \cite{Rupperiner1996}. The number density of the speculative black hole molecules was introduced to examine its phase transition and microstructure. This method was quickly generalized to other black holes \cite{Ruppeiner2018,Miao2019}. Furthermore from the view point of the Gibbs free energy landscape, the dynamic properties of various black hole phase transitions were studied in Refs. \cite{Li2020,Wei2021,Yang2021,Kumara2021,Mo2021}. Since the phase transition of black holes is more important in black hole thermodynamics, there exists a natural question how to probe the information of the black hole phase transition.

So far although experimental evidences of black hole thermodynamics have not been found, people still expect that black hole thermodynamics should have some observational signatures, for example the quasinormal modes (QNMs), which may be detected from astronomical observations in the future. Recently, the authors in Refs. \cite{Jing2008,Berti2008,He2008,Shen2007,Rao2007,Liu2014,Mahapatra2016,Chabab2016,Prasia2017,Zou2017,Liang2019} had found that QNMs indeed exhibit characteristic behaviours along the black hole phase transition. Thus QNMs can be use to probe the phase transition of black holes in AdS/dS spacetime. Especially there is a dramatic change of the slope of QNMs along with the black hole phase transition \cite{Liu2014}. These results indicate that there is a close connection between the dynamical and thermodynamical aspects of black holes. In addition, as we well known that QNMs are intimately linked to the unstable photon orbits \cite{Goebel1972,Ferrari1984,Mashhoon1985}, then they are playing an important role in strong gravitational phenomena (such as the shadow, lensing, and gravitational waves) \cite{Bozza2002,Cardoso2009,Stefanov2010,Wei2011,Hod2011,Wei2014,Raffaelli2016,Cardoso2016,Konoplya2017}. In short, QNMs can be regarded as the vibration frequencies of photon orbits and can be derived by the photon orbits \cite{Dolan2009}. Based on these observations, it is natural to construct the relation between the photon orbits and black hole phase transitions and to check out whether properties of photon orbits can signal the black hole phase transition.

Firstly, people proposed that the photon orbits signals a possible York-Hawking-Page phase transition in Refs. \cite{Cvetic2016,Gibbons2008}. Then the more clearly relation between photon orbits and phase transitions of Reissner-Nordstrom-AdS black hole and rotating Kerr-AdS black hole were shown in Ref. \cite{Wei2018}. Namely, the photon orbit radius and the minimum impact parameter along with the temperature are both exhibiting non-monotonic behaviours for the low pressure, whose behaviours are indeed the characters of the phase transition existence. Particularly, the changes of those two quantities near the critical point could be treated as the order parameters of phase transition, and they have an universal exponent $1/2$. Subsequently, this issue has been extended to other black holes \cite{Kynara2020} and gravity theory (Gauss-Bonnet case \cite{Han2018}, massive gravity case \cite{Chabab2019}, Born-Infeld case \cite{Xu2019,Li2020NPB}). A general proof of the existence for these relations between photon orbits and phase transitions were shown in Ref. \cite{Zhang2019}.

The linear charged black holes in AdS spacetime \cite{Chamblin1999} within a second-order phase transition show a scaling symmetry: at the critical point the state parameters scale respects to charge q, i.e., $S\sim q^2,~P\sim q^{-2},~T\sim q^{-1}$ \cite{Johnson2018}. It is naturally to gauss whether there exists the scaling symmetry in the non-linear charged AdS black holes. As a generalization of the charged AdS Einstein-Maxwell black holes, it is interesting to explore new non-linear charged systems. Due to infinite self-energy of point like charges in Maxwell's theory \cite{Born1934,Kats2007,Anninos2009,Cai2008,Seiberg1999}, Born and Infeld proposed a generalization when the field is strong, bringing in non-linearities \cite{Dirac2013,Birula1970}. An interesting non-linear generalization of charged black holes involves a Yang-Mill field exponentially coupled to Einstein gravity, which possesses the conformal invariance and is easy to construct the analogues of the four-dimensional Reissner-Nordstr\"{o}m black hole solutions in higher dimensions. Additionally several features of the Einstein-power-Yang-Mills (EPYM) gravity in extended thermodynamics have recently been studied \cite{Du2021,Zhang2015,Moumni2018}.

Inspired by these, we mainly investigate the relation between photon orbits and phase transitions of the charged EPYM AdS black hole. This work is organized as follows. In Sec. \ref{scheme2}, we would like to briefly review the thermodynamic quantities and the second-order phase transition of charged EPYM AdS black hole. The mull geodesics of this system and the deflection angle are also analyzed. In Sec. \ref{scheme3}, we present the behaviour of the photon orbit radius and minimum impact parameter with the temperature nearby the phase transition point and explore the influence of non-linear YM charge parameter on the phase transition. A universal exponent for the changes of photon orbit radius and minimum impact parameter is also obtained. A brief summary is given in Sec. \ref{scheme4}.

\section{Null geodesics of Non-Linear charged AdS black holes}
\label{scheme2}
The action for four-dimensional Einstein-power-Yang-Mills (EPYM) gravity with a cosmological constant $\Lambda$ was given by \cite{Zhang2015,Corda2011,Mazharimousavi2009,Lorenci2002}
\begin{eqnarray}
I=\frac{1}{2}\int d^4x\sqrt{g}
\left(R-2\Lambda-[Tr(F^{(a)}_{{\mu\nu}}F^{{(a)\mu\nu}})]^\gamma\right)
\end{eqnarray}
with the Yang-Mills (YM) field
\begin{eqnarray}
F_{\mu \nu}^{(a)}=\partial_{\mu} A_{\nu}^{(a)}-\partial_{\nu} A_{\mu}^{(a)}+\frac{1}{2 \xi} C_{(b)(c)}^{(a)} A_{\mu}^{b} A_{\nu}^{c}¡£
\end{eqnarray}
Here, $Tr(F^{(a)}_{\mu\nu}F^{(a)\mu\nu})
=\sum^3_{a=1}F^{(a)}_{\mu\nu}F^{(a)\mu\nu}$, $R$ and $\gamma$ are the scalar curvature and a positive real parameter, respectively; $C_{(b)(c)}^{(a)}$ represents the structure constants of three parameter Lie group $G$; $\xi$ is the coupling constant; and $A_{\mu}^{(a)}$ represents the $SO(3)$ gauge group Yang-Mills (YM) potentials.

For this system, the EPYM black hole solution with the negative cosmological constant $\Lambda$ in the four-dimensional spacetime is obtained by adopting the following metric \cite{Yerra2018}:
\begin{eqnarray}
d s^{2}=-f(r) d t^{2}+f^{-1} d r^{2}+r^{2} d \Omega_{2}^{2},\\
f(r)=1-\frac{2 M}{r}-\frac{\Lambda}{3} r^{2}+\frac{\left(2 q^{2}\right)^{\gamma}}{2(4 \gamma-3) r^{4 \gamma-2}},
\end{eqnarray}
where $d\Omega_{2}^{2}$ is the metric on unit $2$-sphere with volume $4\pi$ and $q$ is the YM charge. Note that this solution is valid for the condition of the non-linear YM charge parameter $\gamma\neq0.75$, and the power YM term holds the weak energy condition (WEC) for $\gamma>0$ \cite{Corda2011}. In the extended phase space, $\Lambda$ was interpreted as the thermodynamic pressure $P=-\frac{\Lambda}{8\pi}$. The black hole event horizon locates at $f(r_+)=0$. The parameter $M$ represents the ADM mass of the black hole and it reads
\begin{eqnarray}
M(S, q, P)=\frac{1}{6}\left[8 \pi P\left(\frac{S}{\pi}\right)^{3 / 2}+3\left(\frac{S}{\pi}\right)^{\frac{3-4 \gamma}{2}} \frac{\left(2 q^{2}\right)^{\gamma}}{8 \gamma-6}+3 \sqrt{\frac{S}{\pi}}\right].\label{M}
\end{eqnarray}
And in our set up it is associated with the enthalpy of the system. The black hole temperature, entropy, and volume were given by \cite{Zhang2015}
\begin{eqnarray}
T=\frac{1}{4 \pi r_{+}}\left(1+8 \pi P r_{+}^{2}-\frac{\left(2 q^{2}\right)^{\gamma}}{2 r_{+}^{(4 \gamma-2)}}\right),~~~~~S=\pi r_{+}^{2},~~~~V=\frac{4\pi r_+^3}{3}.    \label{T}
\end{eqnarray}
The YM potential $\Psi$ was given by \cite{Anninos2009}
\begin{eqnarray}
\Psi=\frac{\partial M}{\partial q^{2 \gamma}}=\frac{r_{+}^{3-4 \gamma} 2^{\gamma-2}}{4\gamma-3}.\label{Psi}
\end{eqnarray}
The above thermodynamic quantities satisfy the first law
\begin{eqnarray}
d M=T d S+\Psi d q^{2 \gamma}+V d P.
\end{eqnarray}
The equation of state $P(V,T)$ for canonical ensemble (fixed YM charge $q$) can be obtained from the expression of temperature as
\begin{eqnarray}
P=\left(\frac{4 \pi}{3 V}\right)^{1 / 3}\left[\frac{T}{2}-\frac{1}{8 \pi}\left(\frac{4 \pi}{3 V}\right)^{1 / 3}+\frac{\left(2 q^{2}\right)^{\gamma}}{16 \pi}\left(\frac{4 \pi}{3 V}\right)^{\frac{1-4\gamma}{3}}\right].\label{PTV}
\end{eqnarray}
It is known that this non-linear AdS black hole exhibits a vdW's-like phase transition, with the critical point given by \cite{Du2021,Du2022}
\begin{eqnarray}
r_{c}^{4 \gamma-2}&=&\left(2 q^{2}\right)^{\gamma} f(1, \gamma), \quad f(1, \gamma)=\gamma(4 \gamma-1),~ S_c=\pi \left(2 q^{2}\right)^{\frac{\gamma}{2\gamma-1}} f^{2}(1, \gamma), \label{Sc}\\
T_{c}&=&\frac{1}{\pi\left(2 q^{2}\right)^{\gamma /(4 \gamma-2)} f^{1 /(4 \gamma-2)}(1, \gamma)} \frac{2 \gamma-1}{4 \gamma-1}, \label{Tc}\\
P_{c}&=&\frac{2 \gamma-1}{16 \pi \gamma\left(2 q^{2}\right)^{\gamma /(2 \gamma-1)} f^{1 /(2 \gamma-1)}(1, \gamma)}.\label{Pc}
\end{eqnarray}
where the YM charge parameter satisfies the condition $\frac{1}{2}<\gamma$. It is clear that this critical point closely depends on the YM charge and the non-linear YM charge parameter. We can obtain a interesting relation, which is only related with the non-linear YM charge parameter, from the above quantities as
\begin{eqnarray}
S_c^2T_c^2P_c=\frac{(2\gamma-1)^3}{16\pi\gamma(4\gamma-1)}.
\end{eqnarray}

\begin{figure}[htp]
\centering
\includegraphics[width=0.45\textwidth]{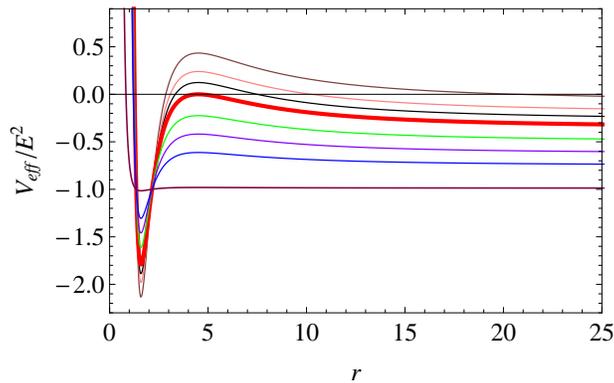}
\caption{The effective potential for the EPYM AdS black h  ole with the parameters $\gamma=1.01,~P=0.003,~q=1.9,~M=2$. The angular momentums $L/E$ of the photon varies from $0.5$ to $36$ from bottom to top. }\label{Veffr}
\end{figure}

Now we will consider a free photon orbiting around a black hole one the equatorial hyperplane defined by $\theta=\pi/2$ and $p_\theta=0$. The Lagrangian takes the form as
\begin{eqnarray}
H=\frac{1}{2}g^{\mu\nu}p_\mu p_\nu=\frac{1}{2}\left(-f^{-1}p_t^2+fp_r^2+r^{-2}p_\phi^2\right),\label{H}
\end{eqnarray}
where the dot represents the derivative to the affine parameter and $p_\mu$ are the generalized momentums. For this black hole background, there are two Killing fields $\partial_t$ and $\partial_\phi$, which lead two constants, the particle energy $E$ and orbital angular momentum $L$ along each geodesics
\begin{eqnarray}
-E=p_t=-f(r)\dot{t},~~~~~~L=p_\phi=r^2\dot{\phi}.
\end{eqnarray}
The light rays are the solutions to Hamilton's equations
\begin{eqnarray}
\dot{p}_\mu=-\frac{\partial H}{\partial x^\mu},~~~~\dot{x}^\mu=\frac{\partial H}{\partial p_\mu},
\end{eqnarray}
which read
\begin{eqnarray}
\dot{p}_t&=&0,~~\dot{p}_\phi=0,~~\dot{p}_\theta=0,\nonumber\\
\dot{p}_r&=&-\frac{1}{2}\left(\frac{f'p_t^2}{f^2}+f'p_r^2+2fp_rp_r'-\frac{2p_\phi^2}{r^3}\right),\nonumber\\
\dot{t}&=&-\frac{p_t}{f},~~\dot{r}=fp_r,~~\dot{\phi}=\frac{p_\phi}{r^2}.\label{dot}
\end{eqnarray}
With $H=0$, we have
\begin{eqnarray}
\dot{r}^2+V_{eff}=0,~~~V_{eff}=\frac{L^2f}{r^2}-E^2.
\end{eqnarray}
As an example, we exhibit the effective potential in Fig. \ref{Veffr} for the EPYM AdS black hole with the fixed parameters $\gamma=1.01,~P=0.003,~q=1.9,~M=2$, and different angular momentums $L/E$. Since the positive of $\dot{r}^2$, there exists the condition: $V_{eff}<0$. Thus the photon can only survive in the range of negative effective potential. When the conserved quantum $L$ is small, the photon will fall into the black hole from somewhere with a larger value of $r$. While, for the larger value of $L$, the peak of effective potential will increase, that leads to the reflection of photon before it falls into black hole. Between these two cases, there exists a critical case which is described by the thickness red line, whose peak approaches zero at $r=4.3122$ and at the same time the radial velocity of photon vanishes. This point just corresponds to the photon sphere because of the spherically symmetric static black hole. In the following, we mainly discuss the relation between the photon sphere radius and phase transition of the EPYM AdS black hole.

\begin{figure}[htp]
\centering
\subfigure[$q=1.9,~\mu_{ph}^2=1,10,21.6,27,28,28.5$]{\includegraphics[width=0.4\textwidth]{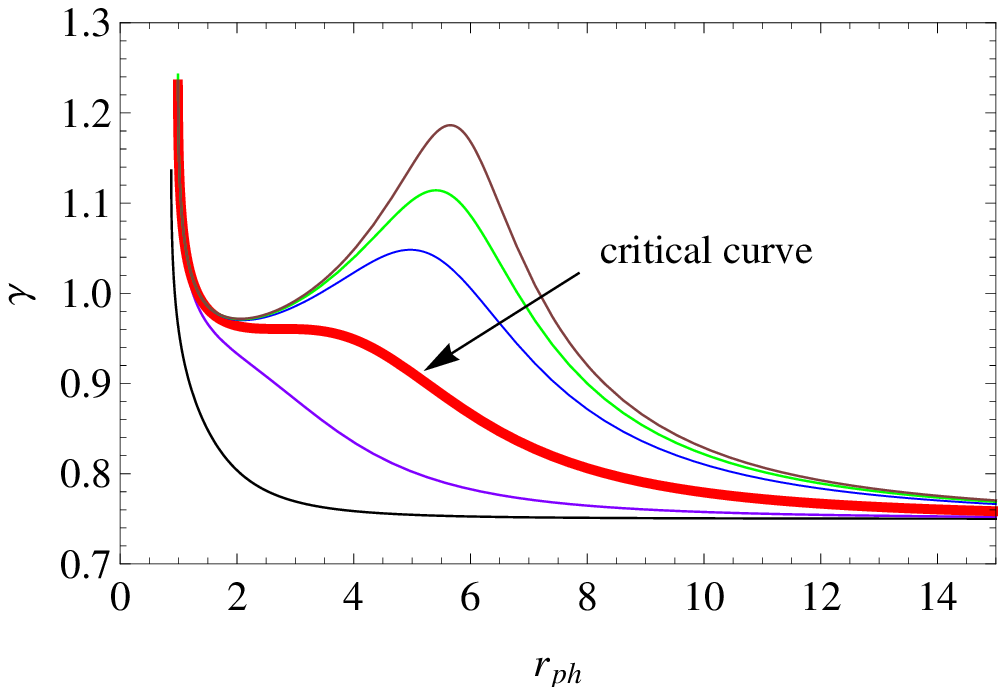}\label{gammarph}}~~~~~~
\subfigure[$q=2,~\gamma=0.85,0.95,0.978,1,1.005,1.01$]{\includegraphics[width=0.4\textwidth]{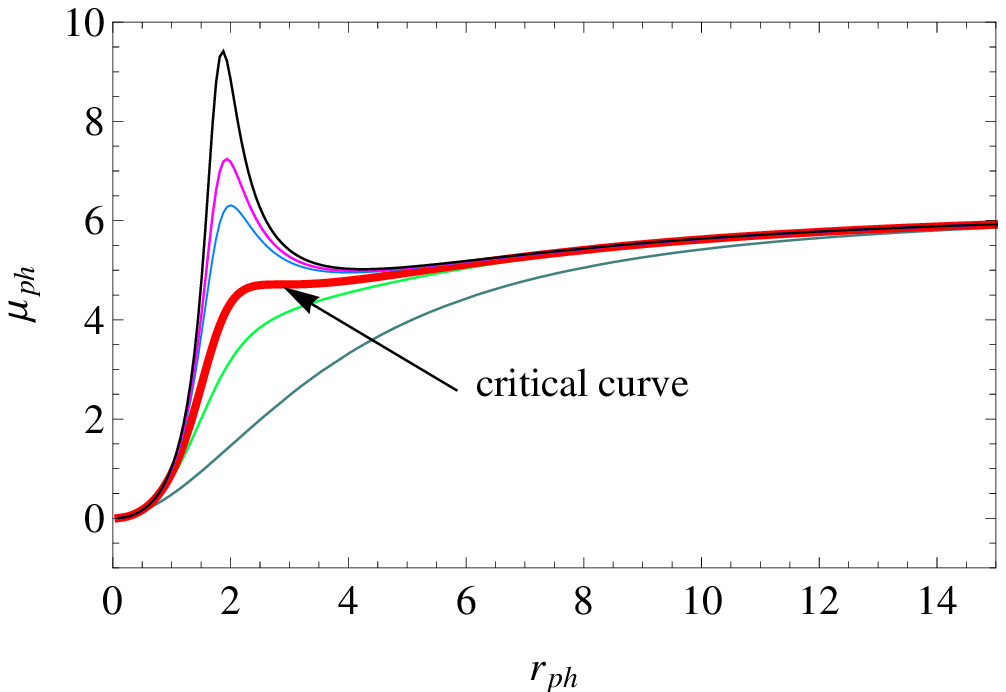}\label{uphrph}}
\caption{$\gamma$ and $\mu_{ph}$ vs. $r_{ph}$ with $P=0.003,~M=2$.}\label{gammauphrph}
\end{figure}

The unstable circular photon sphere is determined by
\begin{eqnarray}
V_{eff}=0,~~\frac{dV_{eff}}{dr}=0,~~\frac{d^2V_{eff}}{dr^2}<0.
\end{eqnarray}
From the second equation, we find that the radius $r_{ph}$ satisfies
\begin{eqnarray}
2f(r)\mid_{r_{ph}}=rf'(r)\mid_{r_{ph}}.\label{rph}
\end{eqnarray}
For given the metric function, we also obtain the radius from the above equation. Solving the first equation, the impact parameter or the angular momentum of the photon is
\begin{eqnarray}
\mu_{ph}\equiv\frac{L}{E_c}=\frac{r}{\sqrt{f(r)}}\bigg|_{r_{ph}}.
\end{eqnarray}
From this relation, we have the behaviors of the non-linear YM charge parameter and the impact parameter with the photon sphere radius, which are shown in Fig. \ref{gammauphrph}.
\begin{figure}[htp]
\centering
\includegraphics[width=0.45\textwidth]{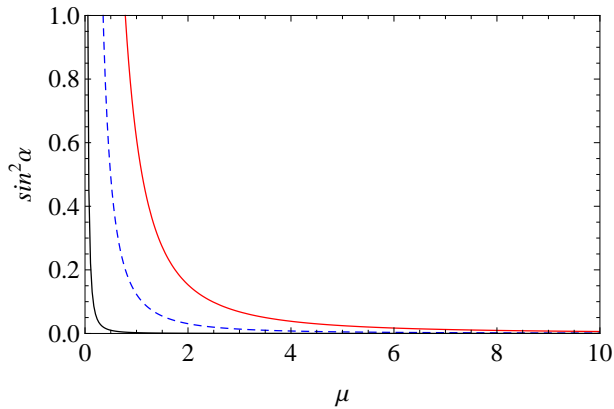}
\caption{The angle as the function of impact parameter with the parameters $\gamma=1,~P=0.0003,~q=1.9,~M=2$. The observer's  position varies from $1$ to $1000$ from the red line to the black one. }\label{sin2u}
\end{figure}
It is obviously that there exist the critical curves when given the pressure and mass of black hole. The corresponding critical values are exhibited in Table. \ref{critical}. The critical impact parameter, radius, and non-linear YM charge parameter are both increasing with the YM charge.
\begin{table}[htbp]
\centering
\caption{The critical quantities in $r_{ph}-\gamma$ diagram for different YM charges {\color{red}and $M=2$}.}
\begin{tabular}{|c|c|c|c|c|c|}
\hline
\centering
$q$~~&$0.85$~~& $1$~~&$1.5$~~&$1.9$~~&$2$~~ \\ \hline
~~$\mu_{phc}^2$ ~~&7.351515~~&10.18365~~&17.91095~~&21.5645~~&22.216~~\\ \hline
$r_{phc}$~~&1.06611~~&1.33639~~&2.18652~~&2.74~~&2.86303~~\\ \hline
$\gamma_c$~~&0.804019~~& 0.821639~~& 0.893342~~& 0.960512~~& 0.978177~~\\ \hline
\end{tabular}\label{critical}
\end{table}

Next, we will investigate the effect of the angular momentum on the observer's angle, especially the critical angle with the maximum impact parameter. Considering light rays sent from an observer at radius coordinate $r_0$ into the past. These light rays can be divided into two classes: Light rays of the first class go to infinity after being deflected by the black hole. Light rays of the second one go towards the horizon of the black hole. If there are no light sources between the observer and the black hole, initial directions of the second class correspond to darkness on the observer's sky. This dark circular disk on the observer's sky is called the shadow of the black hole. The boundary of the shadow is determined by the initial directions of light rays that asymptotically spiral towards the outermost photon sphere. Note that the light rays in the photon sphere are unstable with respect to radial perturbations.

We consider a light ray that is sent from the observer's position at $r_0$ into the past under an deflection angle $\alpha$ with respect to the radial direction. The deflection angle is given by
\begin{eqnarray}
\cot{\alpha}=\sqrt{\frac{g_{rr}}{g_{\phi\phi}}}\frac{dr}{d\phi}\bigg|_{r_0}=\sqrt{\frac{1}{f(r)r^2}}\frac{dr}{d\phi}\bigg|_{r_0}.
\end{eqnarray}
From eqs. (\ref{H}) and (\ref{dot}), we have the orbit equation
\begin{eqnarray}
\frac{dr}{d\phi}=\sqrt{f(r)r^2}\sqrt{\frac{r^2 \mu^2}{f(r)}-1}\label{orbit}
\end{eqnarray}
with the impact parameter or the angular momentum $\mu\equiv\frac{L}{E}$. Thus the deflection angle $\alpha$ satisfies
\begin{eqnarray}
\sin^2{\alpha}=\frac{f(r_0)}{r_0^2\mu^2}.
\end{eqnarray}
Especially, the critical angle reads $\sin^2{\alpha_c}=\frac{f(r_0)}{\mu^2_{ph}r_0^2}$. That indicates that $\mu$ has a close relation with the critical deflection angle and the observer's position, which is just the phenomenon of the black hole lensing. For the light ray with the large impact parameter, its deflection angle is small. Gradually decreasing $\mu$ to $\mu_{ph}$, the deflection angle will be larger and larger, until to the bounded one $\alpha_c$. These behaviors are presented in Fig. \ref{sin2u}. That is consistent with that showed in Ref. \cite{Bozza2002}.

As what we have shown in the above, there are two important quantities $r_{ph}$ and $\mu_{ph}$ for a photon sphere. In the following, we will probe the detailed behaviors of these two quantities near phase transition, and discuss whether they carry the phase transition information.

\section{Behaviors of $r_{ph}$ and $\mu_{ph}$ near phase transition point}
\label{scheme3}
\begin{figure}[htp]
\centering
\subfigure[$q=1.9$]{\includegraphics[width=0.4\textwidth]{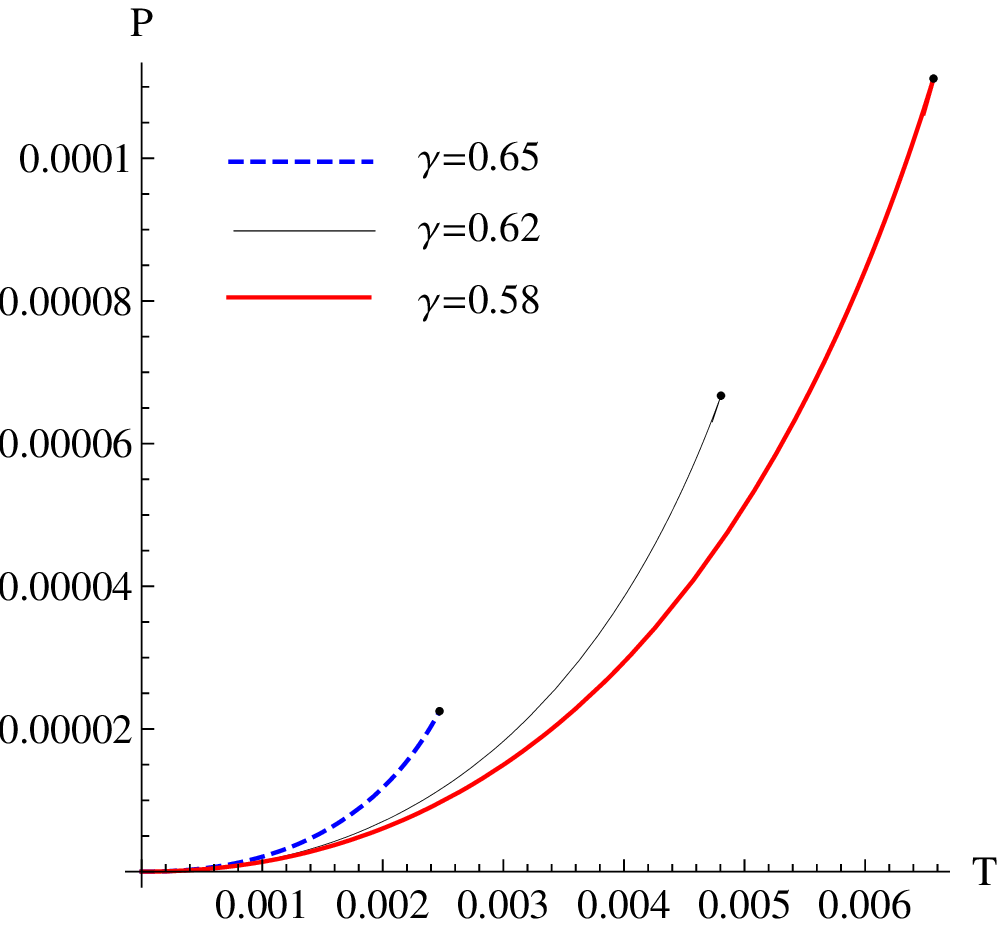}\label{T0-P0-1}}~~~~~~
\subfigure[$q=1.9$]{\includegraphics[width=0.4\textwidth]{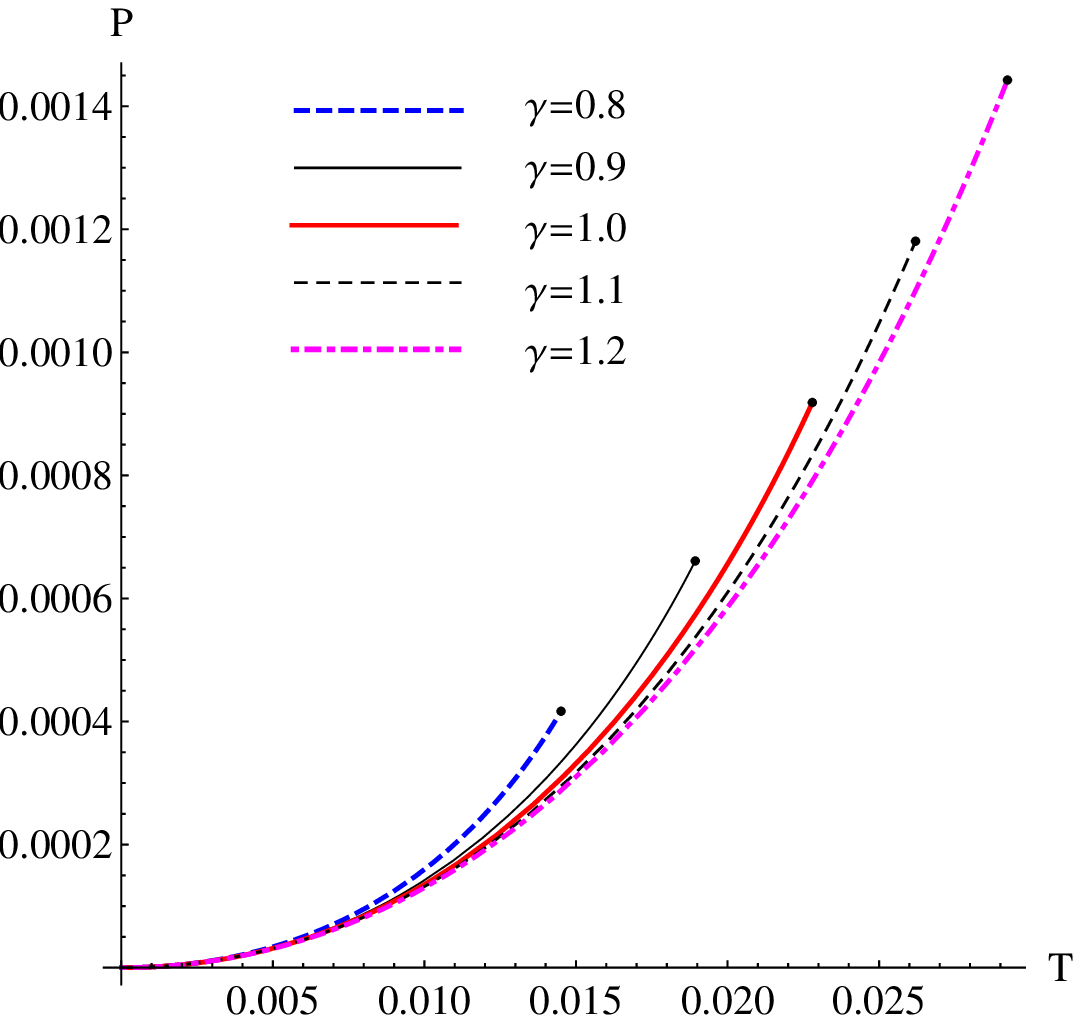}\label{T0-P0-2}}
\caption{The coexistent curve of $P-T$.}\label{T0-P0}
\end{figure}

{\color{red}For the EPYM AdS black hole with the non-linear source, the condition of the first-order phase transition was presented in our work \cite{Du2021} as
\begin{eqnarray}
\chi\frac{2\gamma-1}{\gamma^{1/(4\gamma-2)}(4\gamma-1)^{(4\gamma-1)/(4\gamma-2)}}
=\frac{1}{f^{1/(4\gamma-2)}(x,\gamma)}\left(1+x-
\frac{1-x^{4\gamma}}{2f(x,\gamma)(1-x)x^{4\gamma-2}}\right),~~
\frac{(2q^2)^\gamma}{r_2^{4\gamma-2}}=\frac{1}{f(x,\gamma)} \label{condition}
\end{eqnarray}
with
\begin{eqnarray}
f(x,\gamma)=\frac{(3-4\gamma)(1+x)(1-x^{4\gamma})+8\gamma x^2(1-x^{4\gamma-3})}{2x^{4\gamma-2}(3-4\gamma)(1-x)^3},~~x=\frac{r_{1}}{r_{2}}.
\end{eqnarray}
And we also investigated the effect of the non-linear YM charge parameter $\gamma$ on the phase transition from two different viewpoints. Firstly, from the viewpoint of the coexistent curve ($P-T$), when fixed the YM electric $q$ we found that for a given temperature ($T<T_c$) or a given pressure ($P<P_c$), the pressure or the temperature and the radius of the coexistent big black hole phase increase with the increasing of $\gamma$; while the radius of the coexistent small black hole phase decreases with $\gamma$. Here the two-phases coexistent curves in the $T-P$ plane with different values of the non-linear YM charge parameter $\gamma$ are shown in Fig. \ref{T0-P0}. Secondly, from the viewpoint of the geometry scalar curvature, the absolute values of the scalar curvature for the coexistent big and small black holes in the range of $3/4<\gamma\leq1$ both decrease with the increasing of $\gamma$, while their behavior of the scalar curvature are just opposite in the range of $1\leq\gamma$. In the following, we only exhibit the behaviors of the photon sphere radius as the function of temperature in two cases of $\gamma=1$ (the analytical solution) and $\gamma=1.5$ (the numerical Solution), respectively. Since the solutions of the photon sphere radius with other values of $\gamma$ are difficult to present, so we cannot give the relevant analysis and conclusions.}

\begin{figure}[htp]
\centering
\subfigure[$r_{ph}-T~~\gamma=1$]{\includegraphics[width=0.4\textwidth]{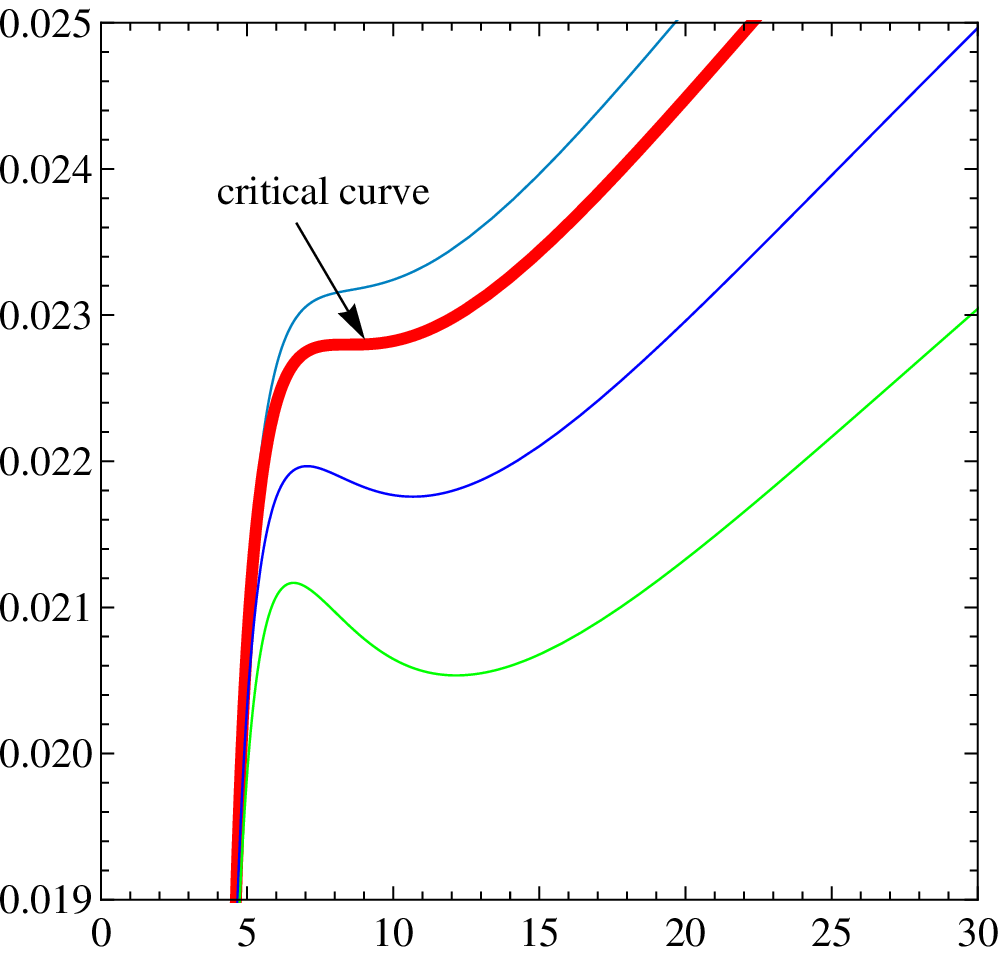}\label{Trph}}~~~~~~
\subfigure[$\mu_{ph}^2-T~~\gamma=1$]{\includegraphics[width=0.4\textwidth]{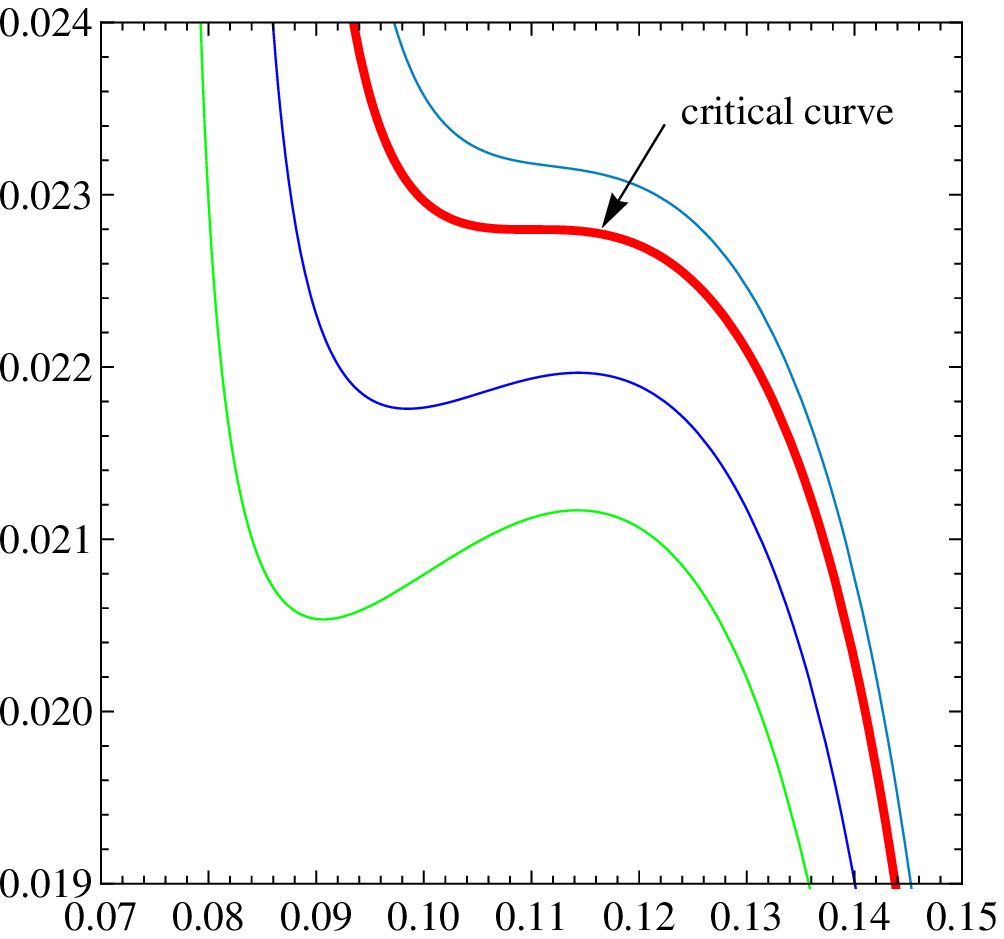}\label{Tuph}}
\caption{The YM charge set to $q=1.9$. The pressure varies from $P_c-0.0002$ to $P_c+0.00004$ from bottom to top. }
\end{figure}

{\color{red}For $\gamma=1$, the photon sphere radius of this system} can be expressed as
\begin{eqnarray}
r_{ph}=\frac{1}{2}\left(3M+\sqrt{9M^2-8q^2}\right).
\end{eqnarray}
It is clear that this result is exactly equal to that of the asymptotically flat charged black hole, in which the orbit radius depends on the pressure. However, we need to note that the mass of eq. (\ref{M}) is related with the pressure. Substituting it into above equation, we have the photon sphere radius
\begin{eqnarray}
r_{ph}=\frac{3\pi q^2+S(3+8P S)+\sqrt{[3\pi q^2+S(3+8P S)]^2-32\pi q^2S}}{4\sqrt{\pi S}}.\label{rph1}
\end{eqnarray}
At the critical point given in eqs. (\ref{Sc}) and (\ref{Pc}), the critical radius reads
\begin{eqnarray}
r_{phc}=(2+\sqrt{6})q.
\end{eqnarray}
The temperature in eq. (\ref{T}) can be rewritten as
\begin{eqnarray}
T(S,P,q)=\frac{1}{4\sqrt{\pi S}}\left(1+8P S-\frac{\pi q^{2}}{S}\right).\label{TSPq}
\end{eqnarray}
Combining eqs. (\ref{rph1}) and (\ref{TSPq}), we give the behavior of temperature as a function of the photon sphere radius with the fixed pressure, which is shown in Fig. \ref{Trph}. It is obviously that for $P<P_c$, there will be a non-monotonic behavior. However, the temperature will become only a monotone increasing function of the photon sphere radius when the pressure is bigger than the critical one. At the critical pressure, there is a deflection point at $r_{ph}=r_{phc}$. That behavior is very similar to the isobar of the vdW's system in $T-S$ plane, which means there exists the first-order phase transition. Thought constructing the Maxwell's equal-area law, we also obtain the phase transition point.

\begin{figure}[htp]
\centering
\subfigure[$\gamma=1$]{\includegraphics[width=0.4\textwidth]{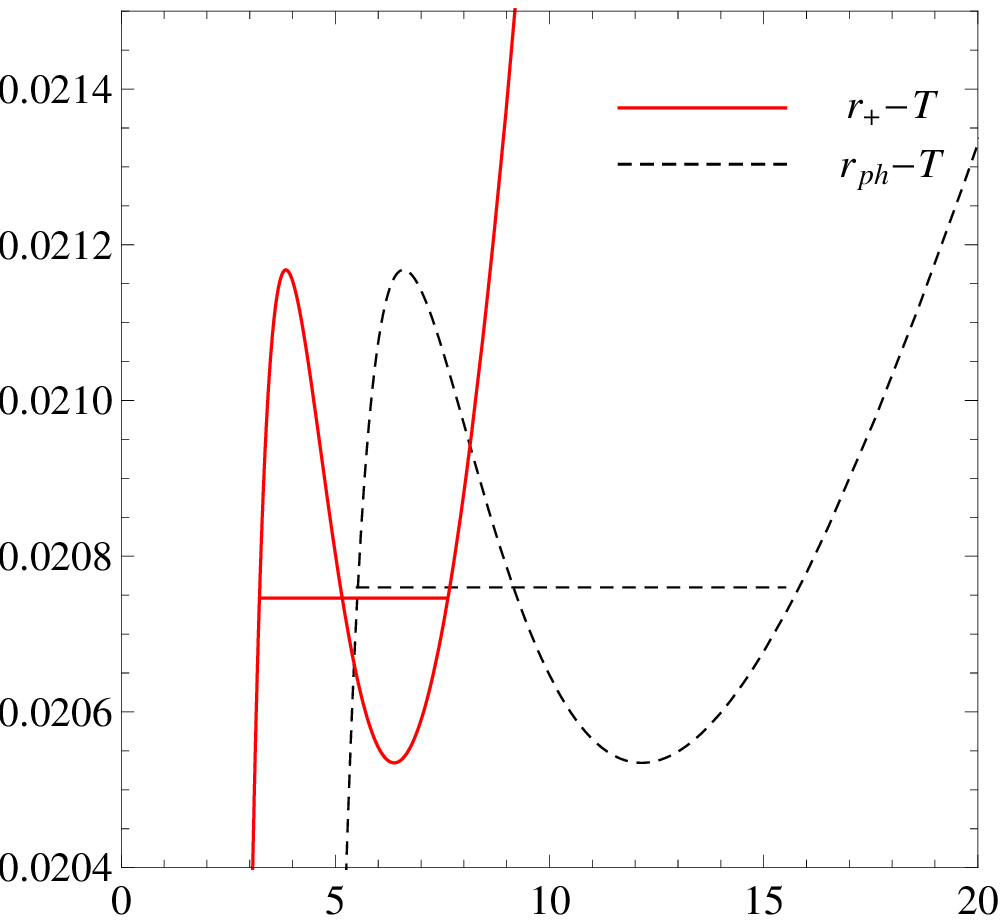}\label{Trphr1}}~~~~~~
\subfigure[$\gamma=1.5$]{\includegraphics[width=0.4\textwidth]{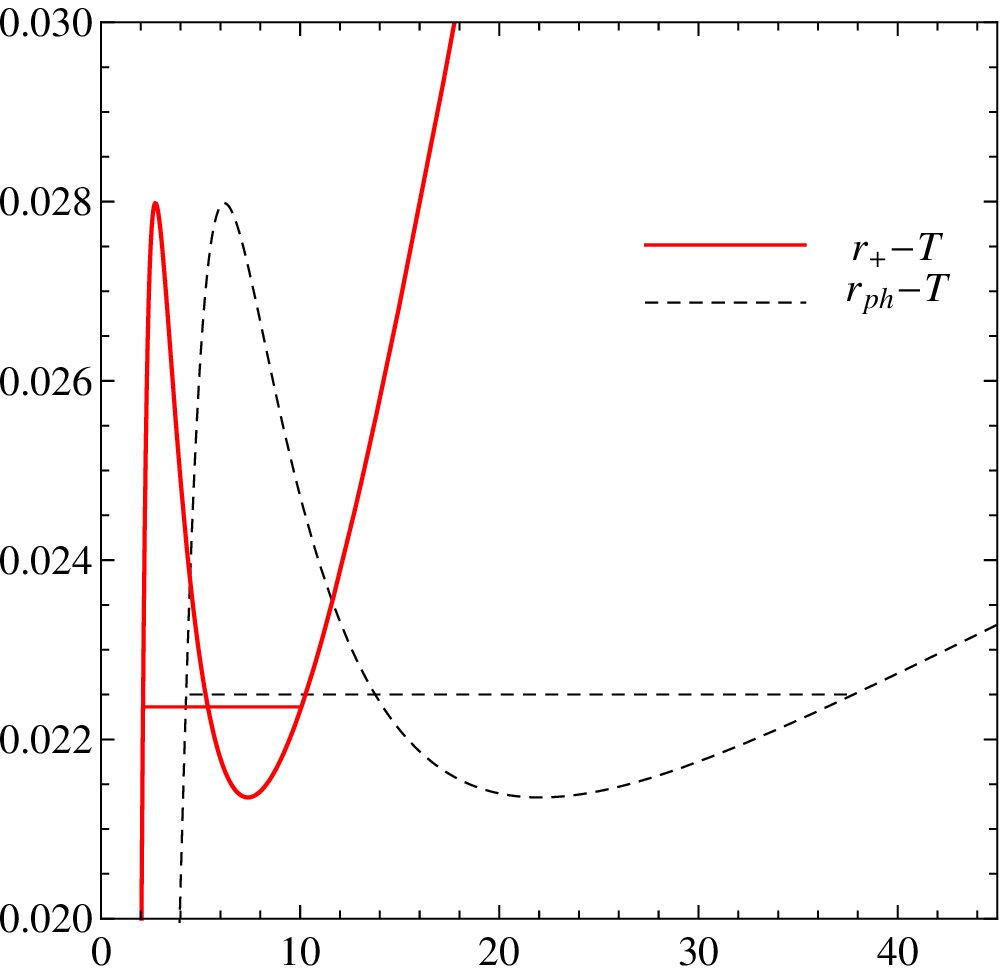}\label{Trphr2}}
\caption{The parameters set to $q=1.9$ and $P=P_c-0.0002$.}
\end{figure}

The behavior of the temperature as a function of the impact parameter square in Figs \ref{Tuph}. Interestingly, it also confirms a non-monotonic property for $P<P_c$. And it becomes monotonous for $P\leq P_c$. Note that these non-monotonic behaviors in $r_{ph}-T$ and $\mu_{ph}^2-T$ diagrams only just indicate the existence of the first-order phase transition, but not the first-order one between high potential black hole and low potential black hole. {\color{red}In order to illustrate the phase transition relationship between the radius of the photon sphere and horizon, the phase transition diagrams in the $r_+-T$ and $r_{ph}-T$ planes are exhibited in Fig. \ref{Trphr1}. It is obviously that the phase transition temperature in $r_{ph}-T$ curve is higher than the one in $r_+-T$. And from the viewpoint of the photon sphere radius, the horizon radii corresponding to the big- and small-coexistent black holes are only just extended correspondingly.}

\begin{figure}[htp]
\centering
\subfigure[$r_{ph}-T~~\gamma=1.5$]{\includegraphics[width=0.4\textwidth]{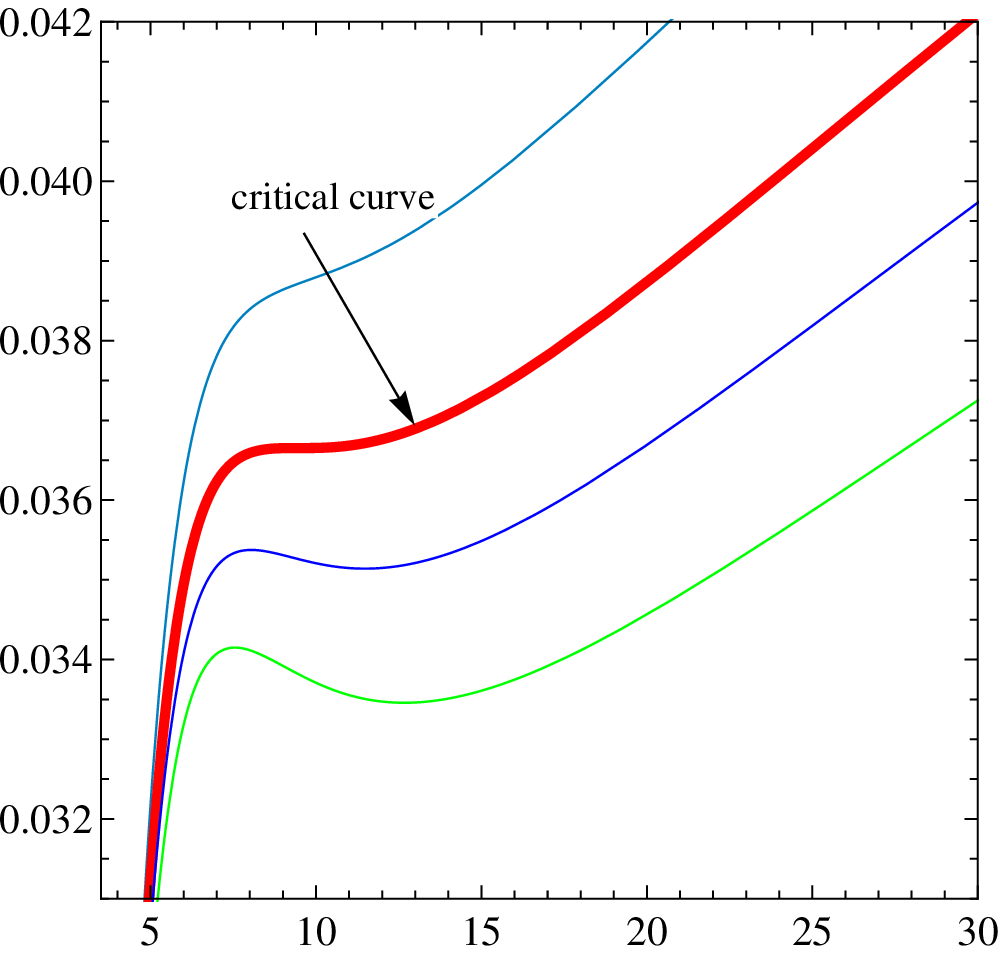}\label{Trph15}}~~~~~~
\subfigure[$\mu_{ph}^2-T~~\gamma=1.5$]{\includegraphics[width=0.4\textwidth]{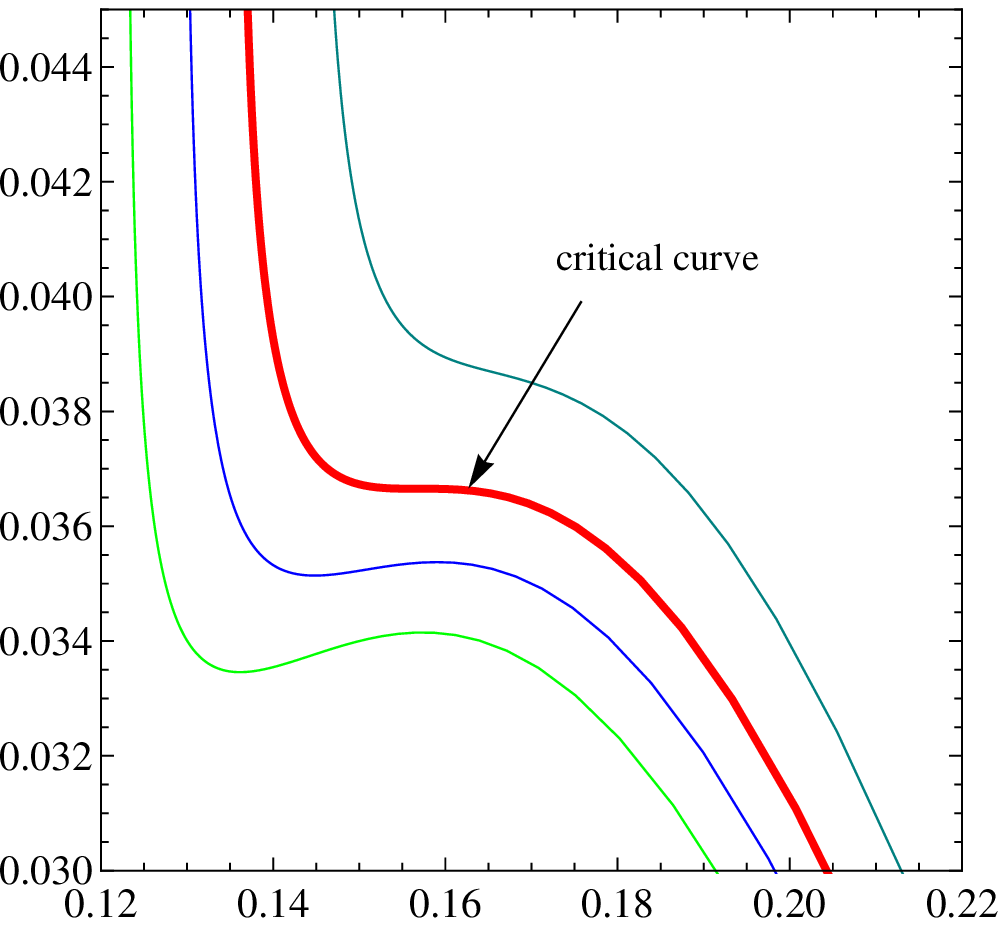}\label{Tuph15}}
\caption{The YM charge set to $q=1.9$. The pressure varies from $P_c-0.0004$ to $P_c+0.00003$ from bottom to top. }\label{Trphuph15}
\end{figure}

\begin{figure}[htp]
\centering
\subfigure[$r_{ph}-T$]{\includegraphics[width=0.4\textwidth]{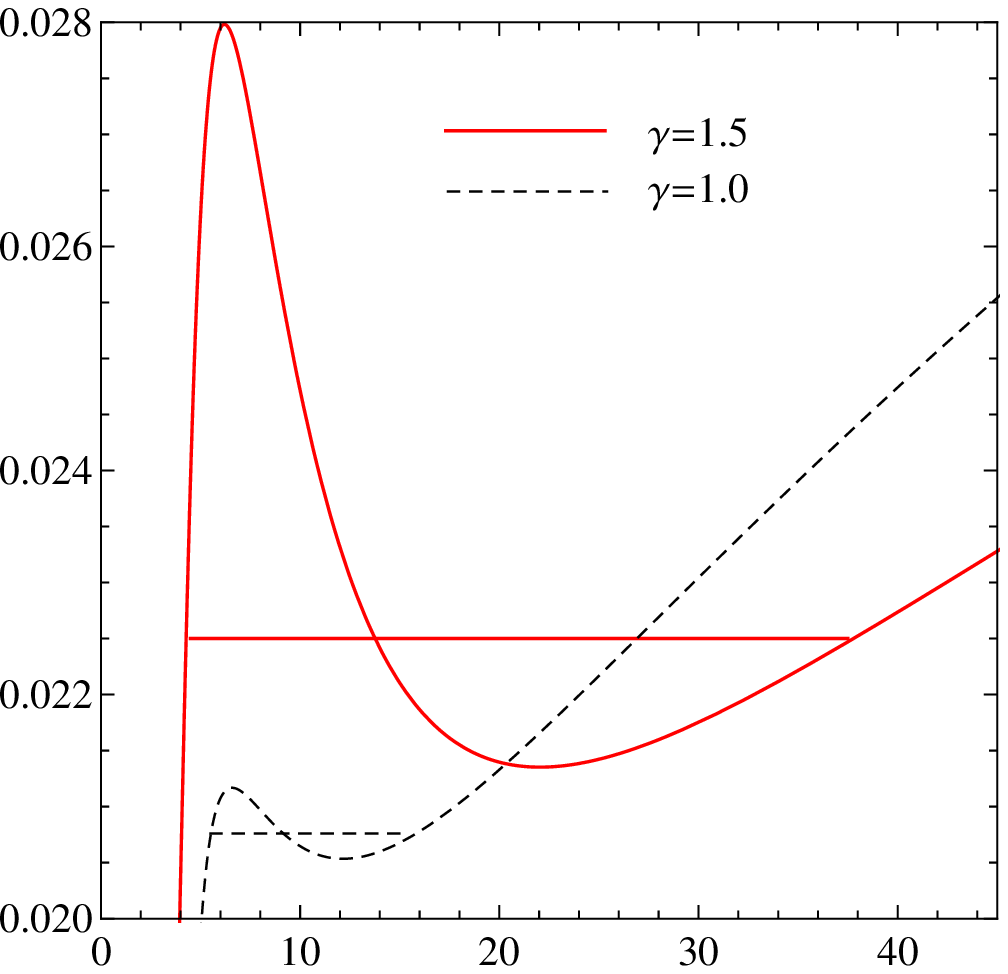}}~~~~~~~~
\subfigure[$\mu_{ph}^2-T$]{\includegraphics[width=0.4\textwidth]{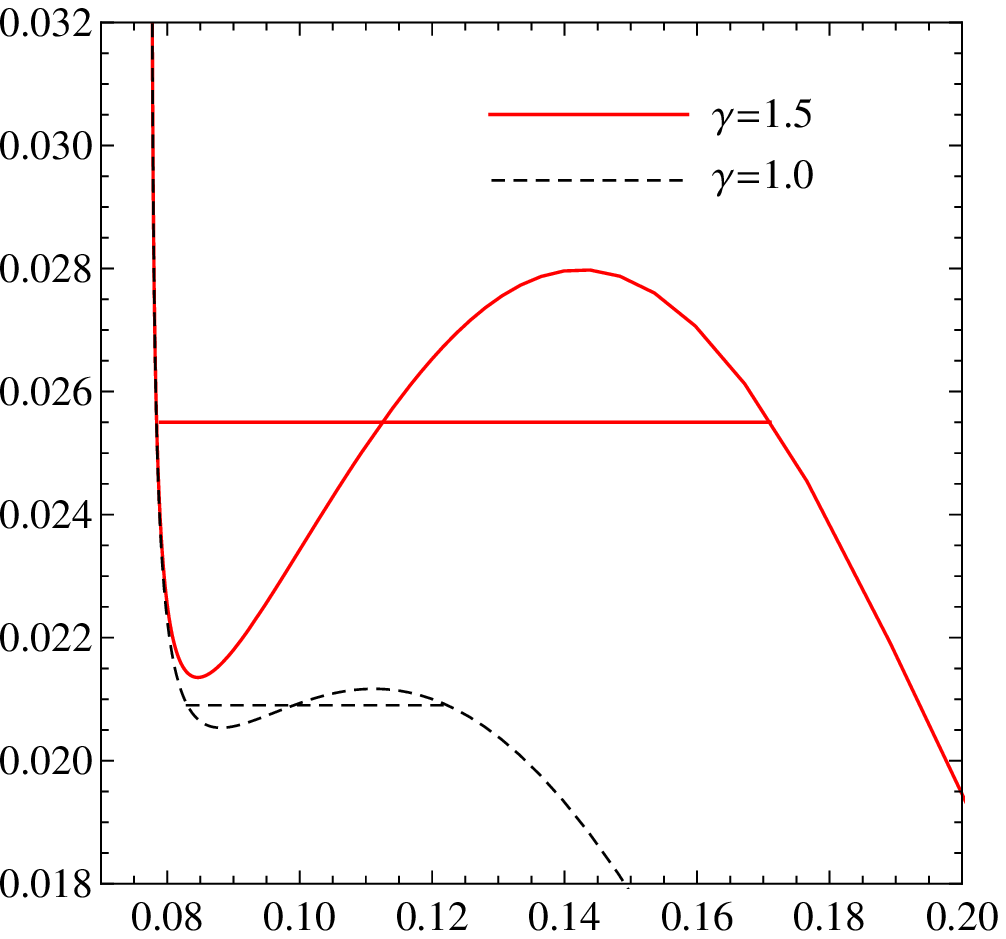}}
\caption{The pressures set to $P_c-0.0002$ and $q=1.9$. }\label{T-rph-P0-addition}
\end{figure}

Furthermore, when $\gamma=1.5$ we also present the numerical results of the temperature with the photon sphere radius and impact parameter for different pressures in Fig. \ref{Trphuph15}. {\color{red}Furthermore, the phase transition relationship from two different viewpoints of the photon sphere radius and the horizon radius is also exhibited in Fig. \ref{Trphr2}. Those behaviors in this case are similar to that in the case of $\gamma=1$. In other words, in both two cases of $\gamma=1$ and $\gamma=1.5$ $r_{ph}$ and $\mu^2_{ph}$ have the non-monotonic behavior and exhibit the first-order phase transition under a certain condition. And from the viewpoint of the photon sphere radius, its radius value and corresponding phase transition temperature are both larger than them from the viewpoint of the horizon radius. These results are the same for this two cases. While the differences shown in Fig. \ref{T-rph-P0-addition} are that: the phase transition temperature with $\gamma=1.5$ is high than the one with $\gamma=1$; the photon sphere radii and impact parameter of the two coexistent phases are different in this two cases.

Note that for the fixed quantities $q$, $\gamma$, and $P$ ($P<P_c$), there also exist the non-monotonic behaviors in the $r_{ph}-T$ and $\mu^2_{ph}-T$ planes; for different points of the $r_{ph}-T$ and $\mu^2_{ph}-T$ curves, the mass parameter of the EPYM AdS black hole is different, which is completely different from the comments discussed in the second part (the mass parameter was fixed). So far, since the behaviors of the photon sphere radius from two different viewpoints have been exhibited, the corresponding results in two parts cannot be directly compared. In the further work, we will focus on the relationship between the observable quantity, the shadow radius and the thermodynamic phase transition of this system in order to realize the experimental detection of the black hole thermodynamic properties when the phase transition occurs.}

\begin{figure}[htp]
\centering
\includegraphics[width=0.4\textwidth]{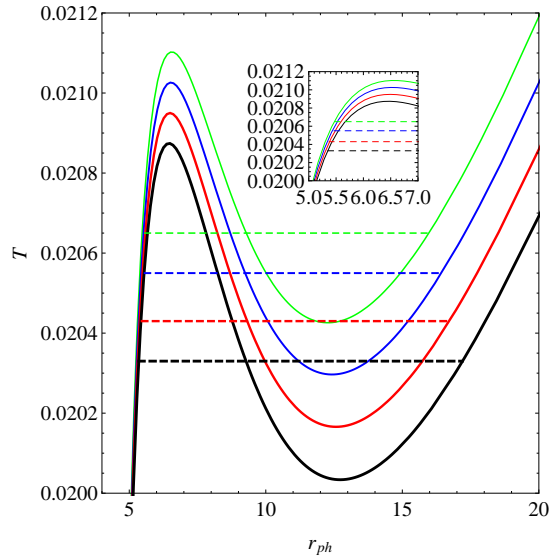}
\caption{{\color{red}The parameters set to $q=1.9$ and $\gamma=1$}. The pressure varies from $P_c-0.00023$ to $P_c-0.0002$ from bottom to top. }\label{EquallawTrph}
\end{figure}

\begin{figure}[htp]
\centering
\subfigure[$\Delta r_{ph}/r_{phc}-T/T_c~~\gamma=1$]{\includegraphics[width=0.4\textwidth]{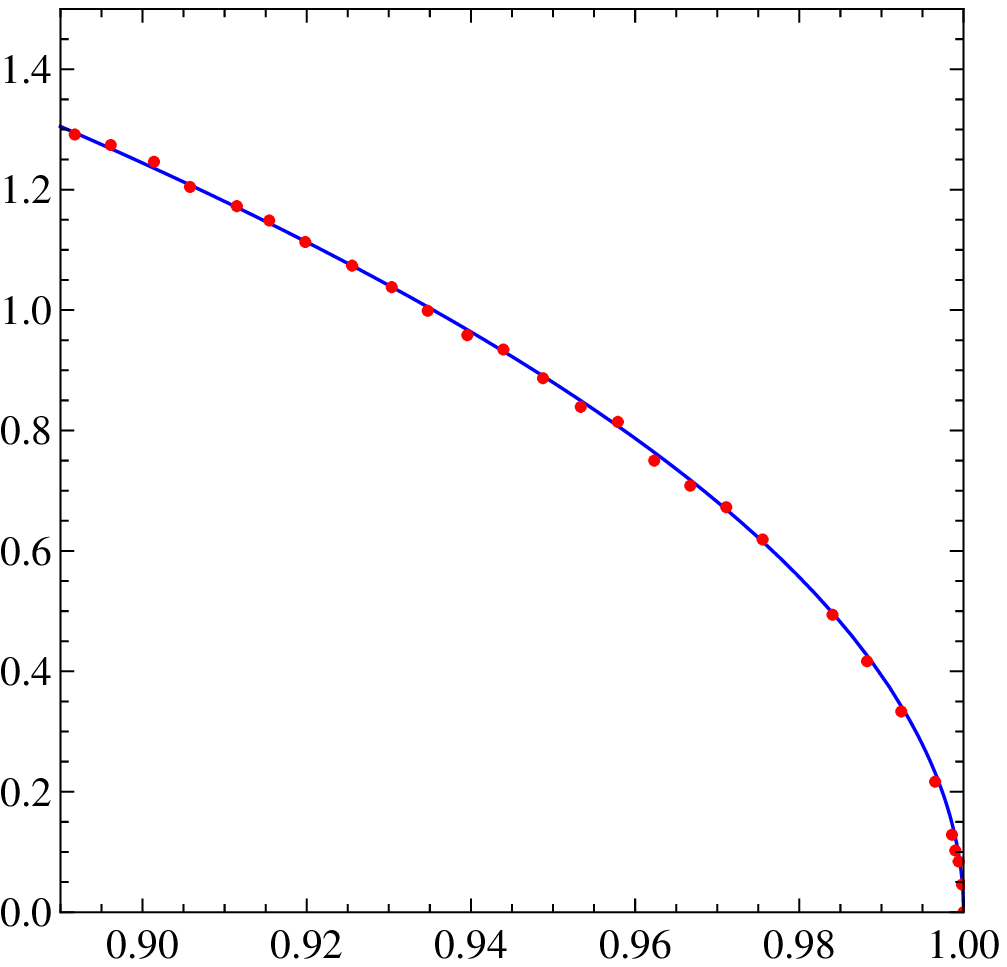}\label{Trph12}}~~~~~~
\subfigure[$\Delta \mu_{ph}/\mu_{phc}-T/T_c~~\gamma=1$]{\includegraphics[width=0.4\textwidth]{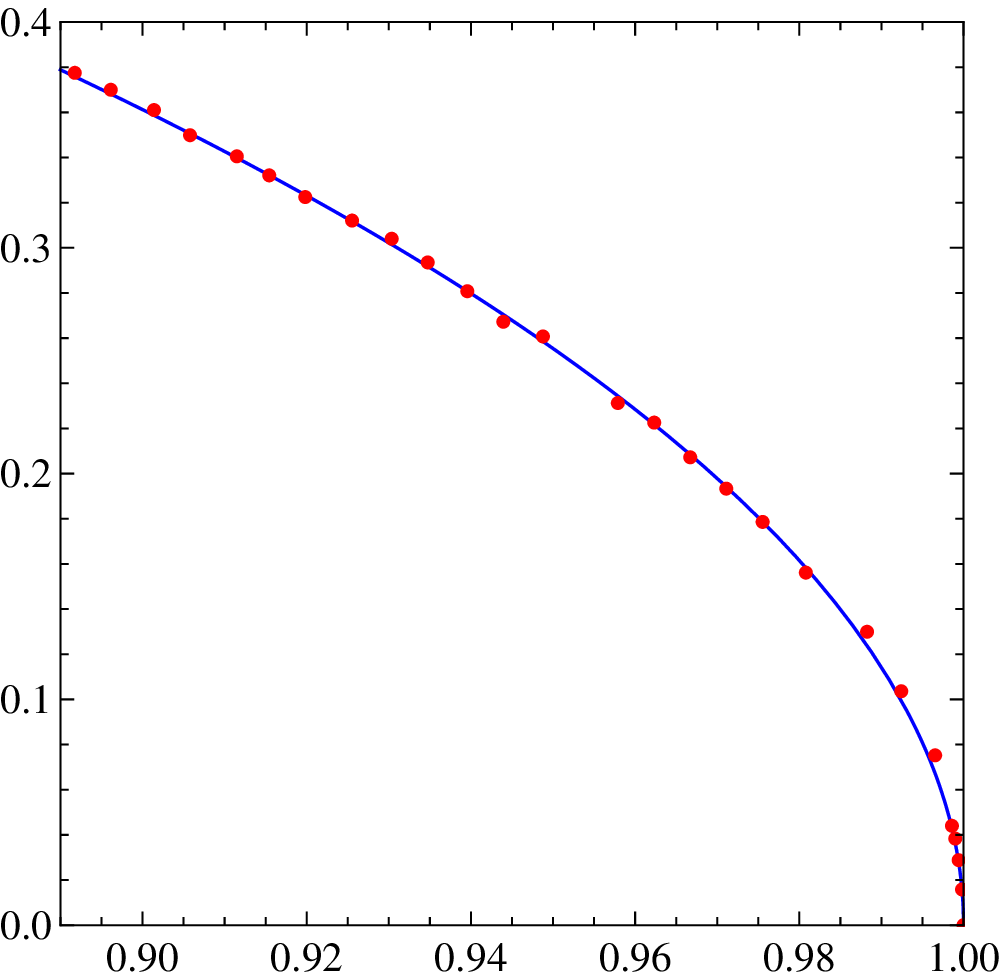}\label{Tuph12}}
\caption{The YM charge set to $q=1.9$. }\label{Trphuph12}
\end{figure}

In order to probe the changes of the photon sphere radius and the impact parameter during the phase transition process, we present the temperature as a function of the photon sphere radius with {\color{red}different pressure ($P<P_c$) in Fig. \ref{EquallawTrph}}. The dashed lines stand for the coexistent phases, the corresponding phase transition temperature is determined by eq. (\ref{TSPq}). With the increasing of photon sphere radius, the phase transitions will emerge at the dashed lines. And there existences the sudden change of photon sphere radius. Further increasing the pressure or temperature to the critical values, the sudden change of photon sphere radius disappears: $\Delta r_{ph}\equiv r_{ph2}-r_{ph1}=0$. Thus if the photon sphere radius has a sudden change, the black hole must be undergoing a first-order phase transition. The similar conclusions hold for the case between the impact parameter and temperature. With the method of numerical drawing, we exhibit the changes of the reduced photon sphere radius and reduced impact parameter ($\Delta r_{ph}/r_{phc}$ and $\Delta\mu_{ph}/\mu_{phc}$) as the functions of the reduced temperature ($T/T_c$) with the parameters $\gamma=1,~q=1.9$ in Fig. \ref{Trphuph12}. It is obviously that both the values of $\Delta r_{ph}/r_{phc}$ and $\Delta\mu_{ph}/\mu_{phc}$ decrease with the reduced temperature, and they are approach to zero until $\frac{T}{T_c}=1$. At the same time the first-order phase transition becomes the second-order one. Therefore the changes of reduced photon sphere radius and reduced impact parameter can be regarded the order parameters for the black hole phase transition.

It is well known that the critical exponent of the order parameter for a ordinary thermodynamic system nearby the critical point equals to $1/2$, whether these two order parameters have the same value of critical exponent. Using the numerical fitting method near $T_c$, we find
\begin{eqnarray}
\Delta r_{ph}/r_{phc}&\sim&3.935\sqrt{1-T/T_c},\\
\Delta\mu_{ph}/\mu_{phc}&\sim&1.1423\sqrt{1-T/T_c}.
\end{eqnarray}
This result exactly confirms that both the critical exponents of $\Delta r_{ph}/r_{phc}$ and $\Delta\mu_{ph}/\mu_{phc}$ equal to $1/2$ at the critical point. That indicates from the viewpoint of thermodynamic black holes have more similar behaviors as the ordinary thermal systems, and they maybe also have the similar microstructures from the perspective of the statistics \cite{Wei2015}. Those conclusions strongly imply there exist a relationship between the photon sphere radius and phase transition of black holes.

\section{Discussions and Conclusions}
\label{scheme4}
In this manuscript, we examined the relationship between the photon sphere radius the thermodynamic phase transition for the four-dimensional charge Einstein-power-Yang-Mills (EPYM) AdS black hole. This will provide a new way to realize the link between the gravity and thermodynamics of black holes.

First, we presented the characters of EPYM AdS black hole phase transition: the critical point depend on the YM charge and non-linear YM charge parameter with the condition $\gamma>1/2$ and $\gamma\neq3/4$. Since the critical point stands for the boundary of the coexistent phases, it is keys to probe the black hole phase structure. Then we investigated the null geodesics of a photon in the equatorial plane of EPYM AdS black hole background. By analyzing the effective potential of photon orbits with the certain parameters, we obtained the photon sphere radius and impact parameter (angular momentum of the photon sphere). The results showed that the non-monotonic behaviors appear in $\gamma-r_{ph}$ and $\mu_{ph}-r_{ph}$ diagrams. That indicated that there exists a certain relationship between the photon sphere radius, impact parameter, and EPYM AdS black hole phase transitions. In addition, we also presented the influence of impact parameter on the reflected angle.

Finally, we further explored the relationship between the unstable photon sphere radius, impact parameter, and thermodynamic phase transition. It mainly contains {\color{red}three aspects as the following:}
\begin{itemize}
  \item{For the isobar and isotherm processes of this system, the photon sphere radius and impact parameter have the non-monotonic behaviors with the certain parameters. When the pressure or temperature is less than the corresponding critical value, these two quantities both have two extremal points. Until the pressure or temperature approaches to the critical one, two extremal points coincide with each other. When the pressure or temperature is more than the critical one, there is no extremal point. Thus, the photon sphere radius and impact parameter are the non-monotonic functions with the pressure or temperature. Those behaviors of the photon sphere radius and impact parameter are consistent with that of the black hole thermodynamics. Therefore, the behavior of photon sphere can be regarded a probe to reveal the thermodynamic phase transition information of black holes;}
  \item{{\color{red}In both two cases of $\gamma=1$ and $\gamma=1.5$, from the viewpoint of the photon sphere radius the corresponding phase transition temperature is higher than that from the viewpoint of the horizon radius. Compared with the horizon radius, the photon sphere radii corresponding to the two coexistent phases are only expanded accordingly; }}
  \item{At the first-order phase transition point, the photon sphere radius and impact parameter both have a sudden change. The changes of them before and after first-order phase transition can act as order parameters to describe EPYM AdS black hole phase transitions. More importantly, at the critical point, the fitting result displays that they have a same critical exponent of $1/2$.}
\end{itemize}

Those analyze firmly enhance our conjecture that there exists the relationship between the null geometry and thermodynamic phase transition for EPYM AdS black holes. It further supports the link between the gravity and thermodynamics of black holes, and also provides a possible way to describe the strong gravitational effect from the thermodynamic side.

\section*{Acknowledgements}

We would like to thank Prof. Ren Zhao, Meng-Sen Ma, and Si-Jiang Yang for their indispensable discussions and comments. This work was supported by the National Natural Science Foundation of China (Grant No. 11705106, 11475107, 12075143), the Natural Science Foundation of Shanxi Province, China (Grant No.201901D111315), the Natural Science Foundation for Young Scientists of Shanxi Province, China (Grant No.201901D211441), and the Scientific Innovation Foundation of the Higher Education Institutions of Shanxi Province (Grant Nos. 2020L0471, 2020L0472, 2016173).

\end{document}